\documentclass[twocolumn]{aastex631}

\usepackage{graphicx}
\usepackage{booktabs}
\usepackage{multirow}
\usepackage{amsmath}

\begin{document}



\title{Exploring Quasar Variability With ZTF at $0<z<3$: A Universal Relation with Eddington Ratio}





\correspondingauthor{Hygor Benati Gonçalves}
\email{hygor.benati@ufrgs.br}

\author[0009-0006-0492-9679]{Hygor Benati Gonçalves}
\affiliation{Departamento de Astronomia, Instituto de f\'isica, Universidade Federal do Rio Grande do Sul, CP 15051, 91501-970, Porto Alegre, RS, Brazil}

\author[0000-0002-5854-7426]{Swayamtrupta Panda}\thanks{Gemini Science Fellow}
\affiliation{International Gemini Observatory/NSF NOIRLab, Casilla 603, La Serena, Chile}
\affiliation{Laborat\'orio Nacional de Astrof\'isica (LNA), Rua dos Estados Unidos 154, Bairro das Na\c c\~oes, CEP 37504-364, Itajub\'a, MG, Brazil}

\author[0000-0003-1772-0023]{Thaisa Storchi Bergmann}
\affiliation{Departamento de Astronomia, Instituto de f\'isica, Universidade Federal do Rio Grande do Sul, CP 15051, 91501-970, Porto Alegre, RS, Brazil} 

\author[0000-0002-8294-9281]{Edward M. Cackett}
\affiliation{Department of Physics and Astronomy, Wayne State University, 666 W. Hancock Street, Detroit, MI 48201, USA}

\author[0000-0002-3719-940X]{Michael Eracleous}
\affiliation{Department of Astronomy and Astrophysics and Institute for Gravitation and the Cosmos, Penn State University, 525 Davey Lab, 251 Pollock Road, University Park, PA 16802}

\begin{abstract}

Quasars, powered by accretion onto supermassive black holes (SMBHs), exhibit significant variability, offering insights into the physics of accretion and the properties of the central engines. In this study, we analyze photometric variability and its correlation with key quasar properties, including black hole mass ($M_{\mathrm{BH}}$) and nuclear luminosities, using 915 quasars with $0\leq z<3.0$ from the AQMES sample monitored within SDSS-V. Variability metrics were derived from approximately 6-year light curves provided by the Zwicky Transient Facility -- ZTF, while SMBH masses and luminosities were obtained from the SDSS DR16 quasar catalog of \citet{wu2022catalog}. We identify a strong anti-correlation between variability amplitude and luminosity, which strengthens with redshift, and a redshift-dependent trend for $M_{\mathrm{BH}}$: a positive correlation at low redshifts, no significant correlation at intermediate redshifts, and an anti-correlation the highest redshifts. Our main finding is a robust anti-correlation between photometric variability amplitude and Eddington ratio, consistent across redshift bins. We present a general equation encapsulating this relationship, that appears to be almost free of redshift dependence, enabling predictions of quasar variability based on accretion parameters or vice-versa. The derived relation with the Eddington ratio provides a unified framework for interpreting variability in active galactic nuclei (AGN) and facilitates future studies of quasar variability using high-cadence surveys, such as the Vera C. Rubin Observatory's Legacy Survey of Space and Time -- LSST.
\end{abstract}


\section{Introduction} \label{sec:intro}

The study of active galactic nuclei (AGN) through long-term monitoring across their broad emission spectrum is well-established in the literature, providing insights into their central structure and physical phenomena associated with the supermassive black hole (SMBH) \citep[for a review, see][]{peterson2001variability}. Quasars, in particular, are known for their significant optical variability. Numerous studies have aimed to correlate this variability with physical properties of the AGN, such as black hole mass ($M_{\text{BH}}$), normalized accretion rate to the Eddington limit ($\lambda_{\text{Edd}}$), and bolometric luminosity ($L_{\text{Bol}}$) over different timescales \citep[see][]{Kozlowski_2017, arevalo2023optical}. The existence of an anti-correlation between luminosity and optical variability in quasars, as well as with the Eddington ratio, has been well-documented for decades \citep{hook1994variability,cristiani1996optical,berk2004ensemble}. This relationship is primarily linked to the black hole accretion rate, of which the Eddington ratio is a key indicator. However, the underlying physics of these relationships is still not well understood, indicating a possible link between variability and the fuel supply process \citep[e.g.][]{wilhite2008variability}.

Meanwhile, the relationship between variability and black hole mass is still debated, with previous studies finding both positive correlations \citep{wold2007dependence,wilhite2008variability,MacLeod2010,lu2019supermassive} and negative correlations \citep{Kelly2009, kelly2013active}, as well as no correlation at all between $M_{\text{BH}}$ and variability \citep{zuo2012correlations,simm2016pan, rakshit2017optical}. Recent studies suggest that these conflicting results could be due to different timescales for measuring variability \citep{arevalo2023optical}, indicating that the correlation only becomes evident when examining variability over short time intervals (30-150 days). However, many of these past studies relied on light curves with short baselines and uneven cadences, potentially biasing the observed correlations - or lack thereof - with black hole physical properties \citep{Kozlowski_2017}. Upcoming long baseline surveys, such as Rubin observatory's Legacy Survey of Space and Time -- LSST \citep{Ivezic_2019, Panda_2019, Czerny_2023} aim to mitigate these issues, with the Zwicky Transient Facility -- ZTF \citep{bellm2014zwicky}  already serving as an important precursor.

In the present study, we use a sample of 915 quasars from the All Quasar Multi-Epoch Spectroscopy-MEDIUM (AQMES-MED) subsample of the ``Black Hole Mapper" project of the Digital Sky Surve V (SDSS-V) \citep{kollmeier2019sdss}, covering the redshift range 0 $\leq z <$ 3, with photometric data from the ZTF. By combining spectral data with time-domain observations, we hope to deepen our understanding of how factors such as black hole mass, bolometric luminosity, and Eddington ratio influence the variability behavior observed in quasars. Moreover, by examining these relationships across a wide range of redshifts, we aim to gain insights into how these correlations evolve with cosmic time, trying to disentangle the effects of the different spectral ranges covered by the ZTF bands at the hosts as a function of redshift.

To investigate these complex relationships effectively, large-scale time-domain surveys are essential for obtaining high-quality light curves that capture quasar variability over different timescales. The ZTF is an invaluable instrument for this purpose, conducting wide-field photometric monitoring with high cadence and sensitivity \citep{bellm2014zwicky}. Operating primarily in three optical bands: ZTF-g, ZTF-r, and ZTF-i, as described in \citet{dekany2020zwicky}, the ZTF allows for detailed tracking of brightness variations, making it ideal for studying quasar variability \citep{Wang_2022, Graham_2022,Neustadt_2023, Wu_2024, Ming_2024}.

The paper is organized as follows. In Section \ref{sec:data}, we describe the sample selection and its characterization. In Section \ref{sec:metho}, we outline the methodology implemented for processing and filtering of the light curves. In Section \ref{sec:results}, we present our results and provide a detailed discussion. Finally, in Section \ref{sec:conclusion}, we summarize our conclusions.

\section{Sample and Data}
\label{sec:data}

\subsection{Sample selection}

As previously mentioned, for this study we use the Type 1 quasar sample from the AQMES program\footnote{\url{https://www.sdss.org/dr18/bhm/programs/aqmes/}}, which is part of the optical time-domain spectroscopic monitoring of quasars under the Black Hole Mapper (BHM) program\footnote{\url{https://www.sdss.org/dr18/bhm/}}, currently being conducted by the SDSS-V survey \citep{kollmeier2019sdss}. The combination of data from the BH reverberation mapping program with previous SDSS data releases (DRs) and other surveys provides temporal coverage ranging from days to two decades. This coverage includes crucial timescales related to accretion, such as the light-crossing time of the inner broad-line region (BLR) and the dynamical timescale of the outer BLR \citep[see, e.g.,][]{lamassa_etal_2015ApJ_timescales, Stern_etal_2018ApJ_timescales, bon2018agn, cackett2021reverberation, panda2024changing}.

\begin{figure*}
        \centering
        \includegraphics[width=\textwidth]{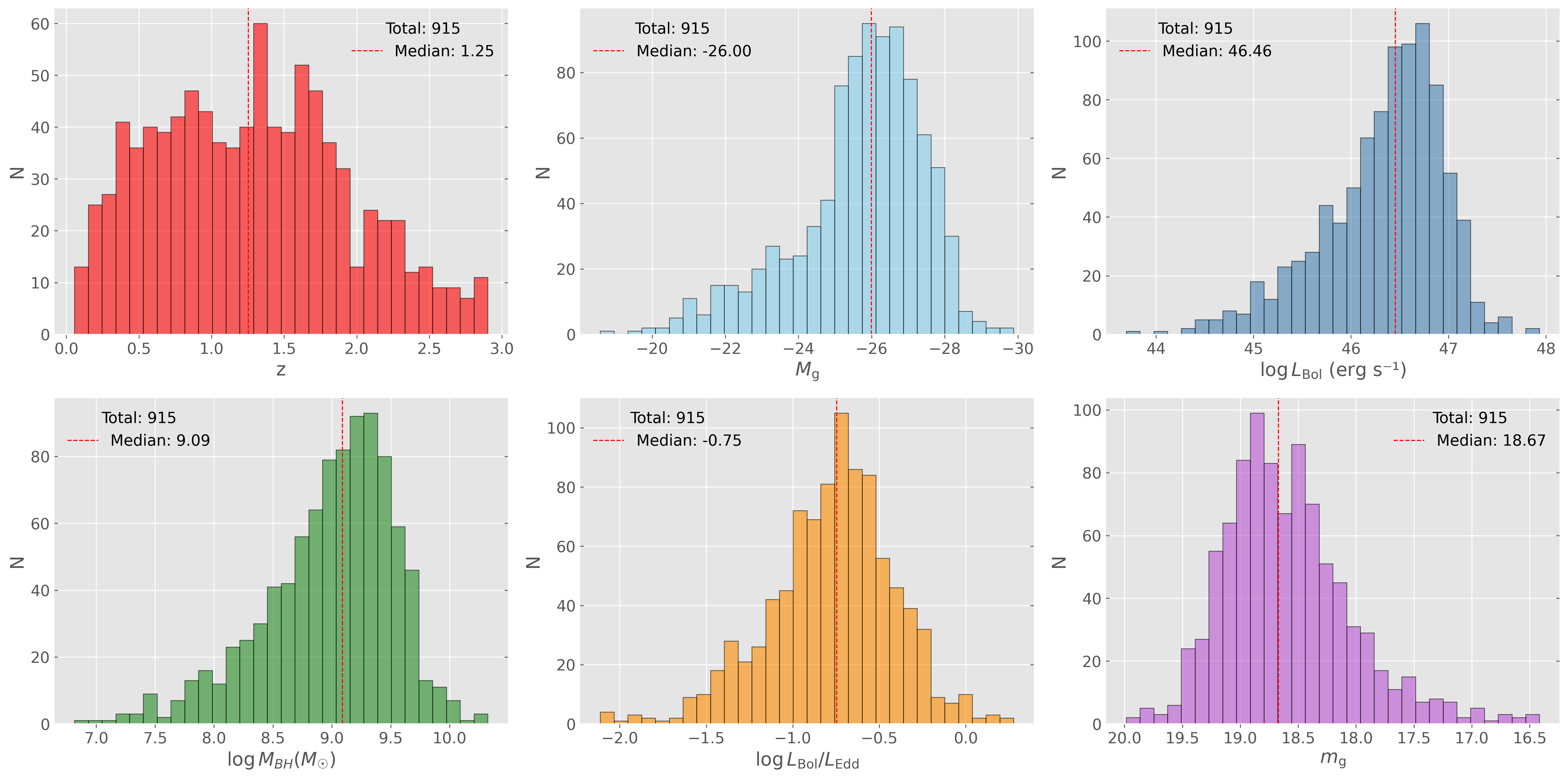}
        \caption{Properties of our filtered sample. \textit{From left to right, top row}: spectroscopic redshift ($z$) from the DR16Q catalog; absolute magnitude in the $g$-band ($M_{\mathrm{g}}$); and bolometric luminosity ($\log L_{\mathrm{Bol}}$, in erg s$^{-1}$). \textit{Bottom row}: black hole mass ($\log M_{\mathrm{BH}}$, in $M_\odot$); Eddington ratio ($\log L_{\mathrm{Bol}}/L_{\mathrm{Edd}}$); and apparent magnitude in the $g$-band ($m_{\mathrm{g}}$). The dashed red line represents the respective median values.}
        \label{combined_histograms}
    \end{figure*}

From the 22,000 quasars in the AQMES sample, we selected the AQMES-MEDIUM carton\footnote{Cartons are groupings of astronomical objects chosen to achieve a common scientific goal.} for photometric variability study. This carton consists of quasars from the SDSS DR16 QSO catalog, observed approximately 10 times over 4 years, and located within 1.49 degrees of at least one AQMES-MEDIUM field center. The uniform observational cadence of this sample is ideal for studying long-term variability and the internal mechanisms of quasars \citep[see, e.g.,][]{cackett2021reverberation, arevalo2023optical, Ricci_2023NatAs_CLAGN_review}. However, it is important to note that this sample is not representative of the broader quasar population. The quasars selected for the SDSS AQMES were chosen based on a variety of specific criteria, resulting in a sample that is preferentially at lower redshifts. This selection bias, as reflected in the distribution shown in Figure \ref{combined_histograms}, suggests that our sample may not fully capture the overall diversity of quasar properties.

After applying the data filtering discussed in the following sections, our final sample consists of 915 quasars with redshifts in the range $0\leq z<3$. We characterized our sample using data from the SDSS DR16 quasar catalog of \citet{wu2022catalog}, which provides key spectroscopic data for each object, including $M_{\text{BH}}$, $L_{\text{Bol}}$, and [OIII]$\lambda$5007 properties. This catalog allows us to extract the spectral features necessary for our analysis. Figure \ref{combined_histograms} displays histograms of the key properties after the filtering.

Table~\ref{tab:mrt_description} summarizes the columns included in the machine-readable table associated with this work. The full version of the table, containing all 915 quasars and their measured properties, is available online\footnote{\url{https://authortools.aas.org/MRT/datafile17266.txt}}.

\begin{deluxetable*}{lll}
\tabletypesize{\scriptsize}
\tablecaption{Description of the columns included in the machine-readable table. \label{tab:mrt_description}}
\tablehead{
\colhead{Label} & \colhead{Units} & \colhead{Description}
}
\startdata
RA & deg & Right ascension (J2000) from AQMES \\
DEC & deg & Declination (J2000) from AQMES \\
SDSS-NAME & --- & SDSS J+[RA][DEC] \\
PLAT & --- & --- \\
MJD & --- & --- \\
FIBERID & --- & --- \\
RA-2 & deg & Right ascension (J2000) from Wu \& Shen DR16 Quasar catalogue (DR16Q) \\
DEC-2 & deg & Declination (J2000) from DR16Q \\
IF-BOSS-SDSS & --- & Source of the input spectrum: BOSS or SDSS \\
Z-DR16Q & --- & Best redshift provided by DR16Q \\
EBV & --- & Milky Way extinction E(B-V) from Schlegel et al. (1998) \\
LOGL5100 & 10$^{-7}$ J s$^{-1}$ & Continuum luminosity at rest frame 5100 Å \\
LOGL5100-ERR & 10$^{-7}$ J s$^{-1}$ & Uncertainty in LOGL5100 \\
OIII5007 & Mixed\tablenotemark{a} & [O III] line properties \\
OIII5007C & Mixed\tablenotemark{a} & [O III] line centroid properties \\
OIII5007-ERR & Mixed\tablenotemark{a} & Uncertainties in [O III] properties \\
OIII5007C-ERR & Mixed\tablenotemark{a} & Uncertainties in [O III] centroid properties \\
LOGLBOL & 10$^{-7}$ J s$^{-1}$ & Bolometric luminosity \\
LOGLBOL-ERR & 10$^{-7}$ J s$^{-1}$ & Uncertainties in bolometric luminosity \\
LOGMBH & $M_\odot$ & Fiducial single-epoch BH mass \\
LOGMBH-ERR & $M_\odot$ & Uncertainty in BH mass \\
LOGLEDD-RATIO & --- & Eddington ratio based on BH mass \\
LOGLEDD-RATIO-ERR & --- & Uncertainty in Eddington ratio \\
Separation & deg & Difference between RA,DEC and RA-2,DEC-2 \\
RMS & mag & G band Root Mean Square deviation \\
Dz & Mpc & Distance luminosity \\
mag-g & mag & Apparent g band magnitude \\
Mag-g & mag & Absolute g band magnitude \\
Var-G & --- & Excess variance in g band \\
Err-G & --- & Error on excess variance \\
\enddata
\tablenotetext{a}{Includes: peak wavelength [0.1 nm], 50\% flux centroid wavelength [0.1 nm], flux [10$^{-17}$ erg s$^{-1}$ cm$^{-2}$], log line luminosity [erg s$^{-1}$], FWHM [km s$^{-1}$], and rest-frame equivalent width [0.1 nm].}
\end{deluxetable*}

\begin{figure}
        \centering
        \includegraphics[width=0.47\textwidth]{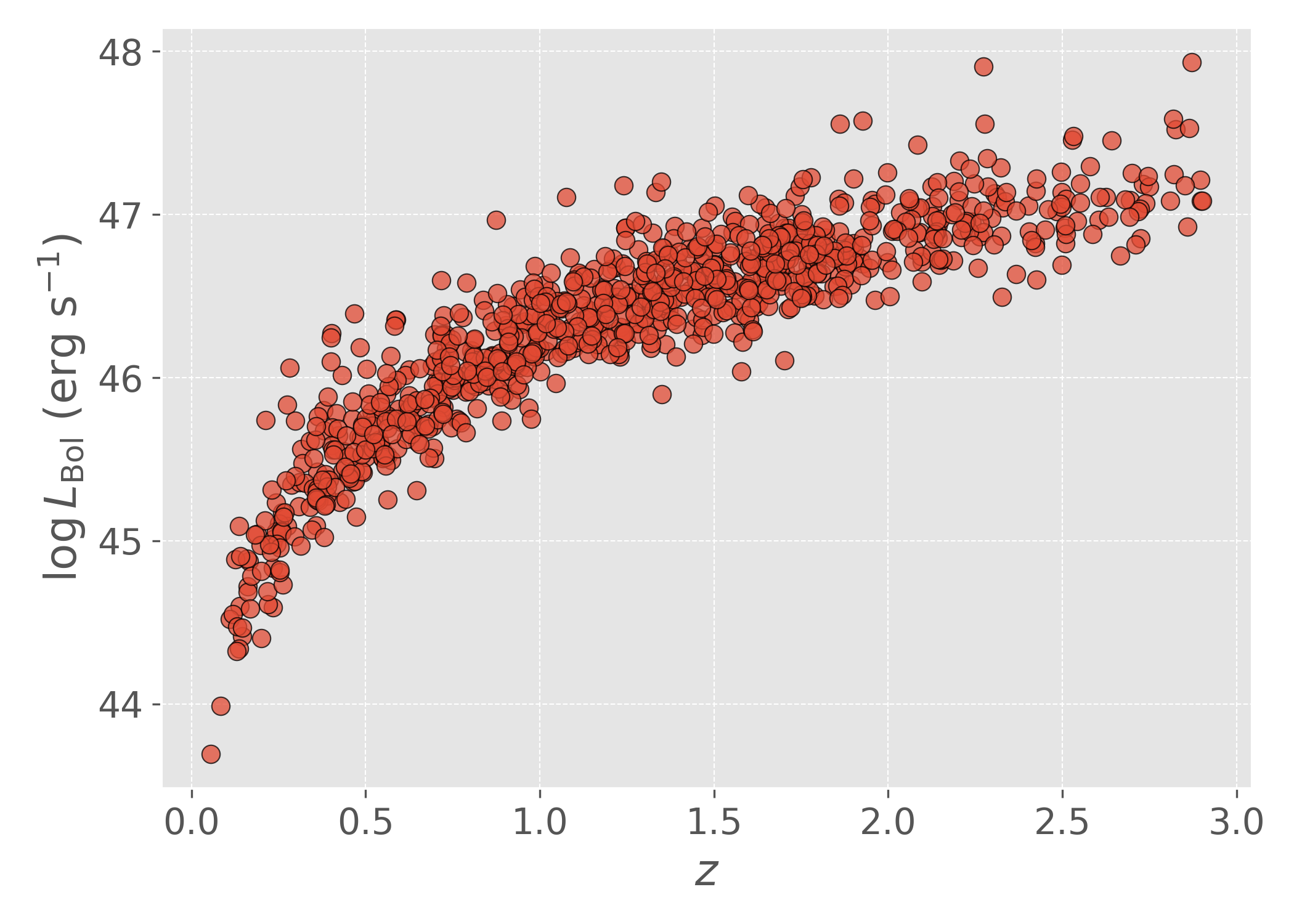}
        \caption{Bolometric luminosity plotted against redshift (z) for the entire filtered sample.}
        \label{lbol_z}
    \end{figure}

We now explore the correlation between bolometric luminosity ($L_{\text{Bol}}$) and redshift ($z$), which highlights selection effects in our study. As shown in Figure \ref{lbol_z}, more luminous AGNs are preferentially observed at higher redshifts, reflecting the well-known effect that the most luminous objects are more easily detected at greater distances. This bias results in a higher observed density of luminous AGNs at high redshifts, which is an inherent feature of optical AGN surveys. 

\subsection{Data}

\begin{figure*}[!ht]
        \centering
        \includegraphics[width=0.95\textwidth]{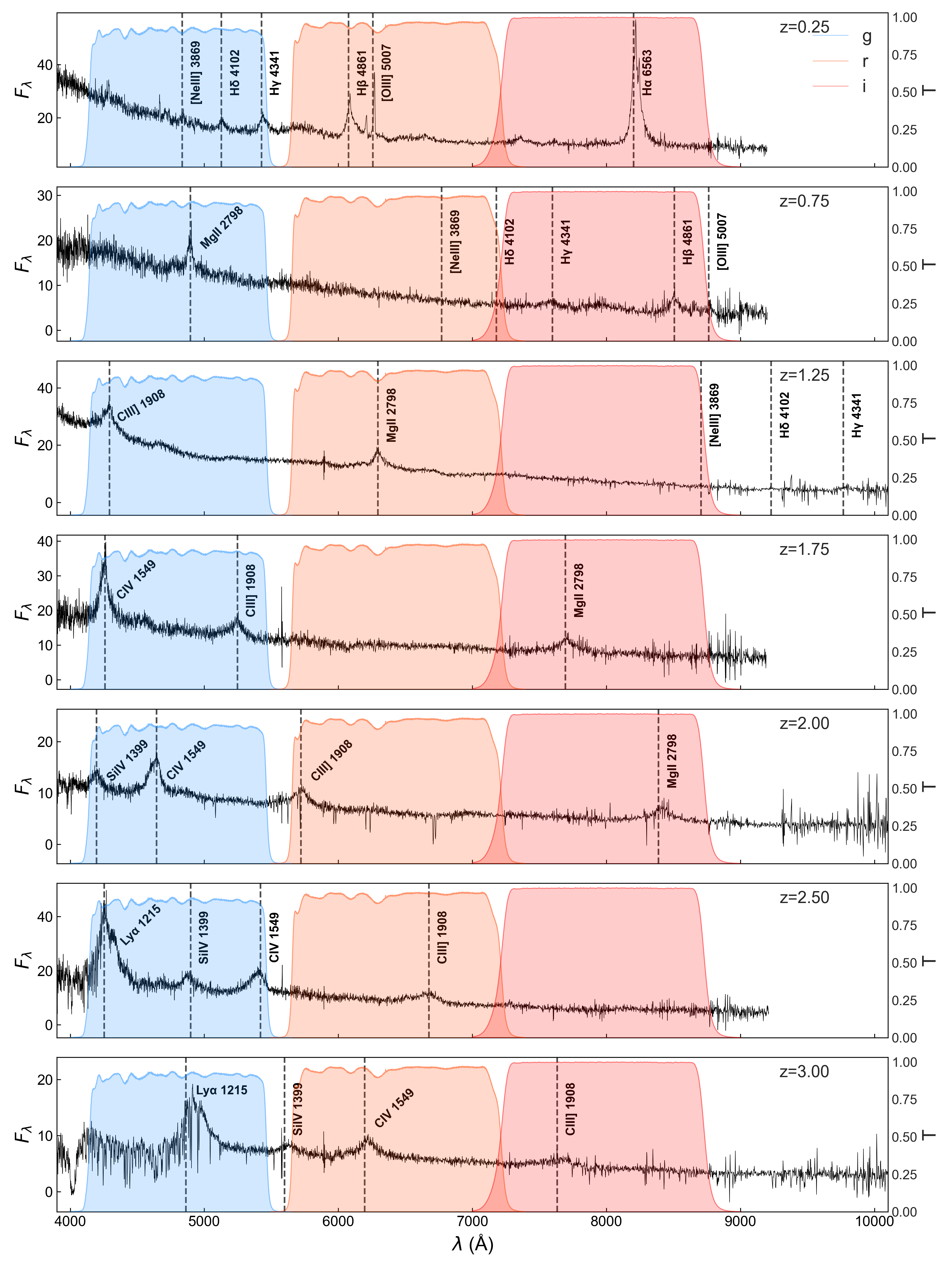}
        \caption{Typical SDSS spectra from our sample, representative of seven redshift bins: 0--0.25, 0.25--1.0, 0.75--1.25, 1.25--1.75, 1.75--2.0, 2.0--2.50, and 2.5--3.0. The ZTF filter transmission profiles are overplotted and typical emission lines are identified. Flux densities ($F_{\lambda}$) are in units of \(10^{-17} \, \text{erg s}^{-1} \text{cm}^{-2} \text{\AA}^{-1}\) and the filters transmission fractions (T) are shown on the right y-axis.}
        \label{fig:ztf+spec}
    \end{figure*}

For the photometric variability analysis, we utilized data from the Zwicky Transient Facility (ZTF) survey, which employs the Samuel Oschin 48" Schmidt telescope at the Palomar Observatory \citep{bellm2018zwicky}. This telescope provides a wide panoramic view spanning 47 square degrees, using a camera equipped with 16 CCDs. The ZTF Science Data System (ZSDS), located at the Infrared Processing and Analysis Center (IPAC) at Caltech, supports ZTF's observational efforts by handling data processing and archiving \citep[for more details, see][] {bellm2018zwicky, masci2018zwicky, graham2019zwicky, dekany2020zwicky}. As mentioned in Sec. \ref{sec:intro}, the ZTF surveys the entire northern sky every three nights using three optical bands: ZTF-g, ZTF-r, and ZTF-i and this regular cadence offers excellent temporal coverage, ideal for studying photometric variability. Figure \ref{fig:ztf+spec} shows typical SDSS spectra from our sample across 7 redshift bins in the observed frame: 0--0.25, 0.25--1.0, 0.75--1.25, 1.25--1.75, 1.75--2.0, 2.0--2.50, and 2.5--3.0. The ZTF filter transmission profiles are overplotted to indicate the spectral region probed by each band at each redshift bin.

\begin{figure*}[!ht]
        \centering
        \includegraphics[width=\textwidth]{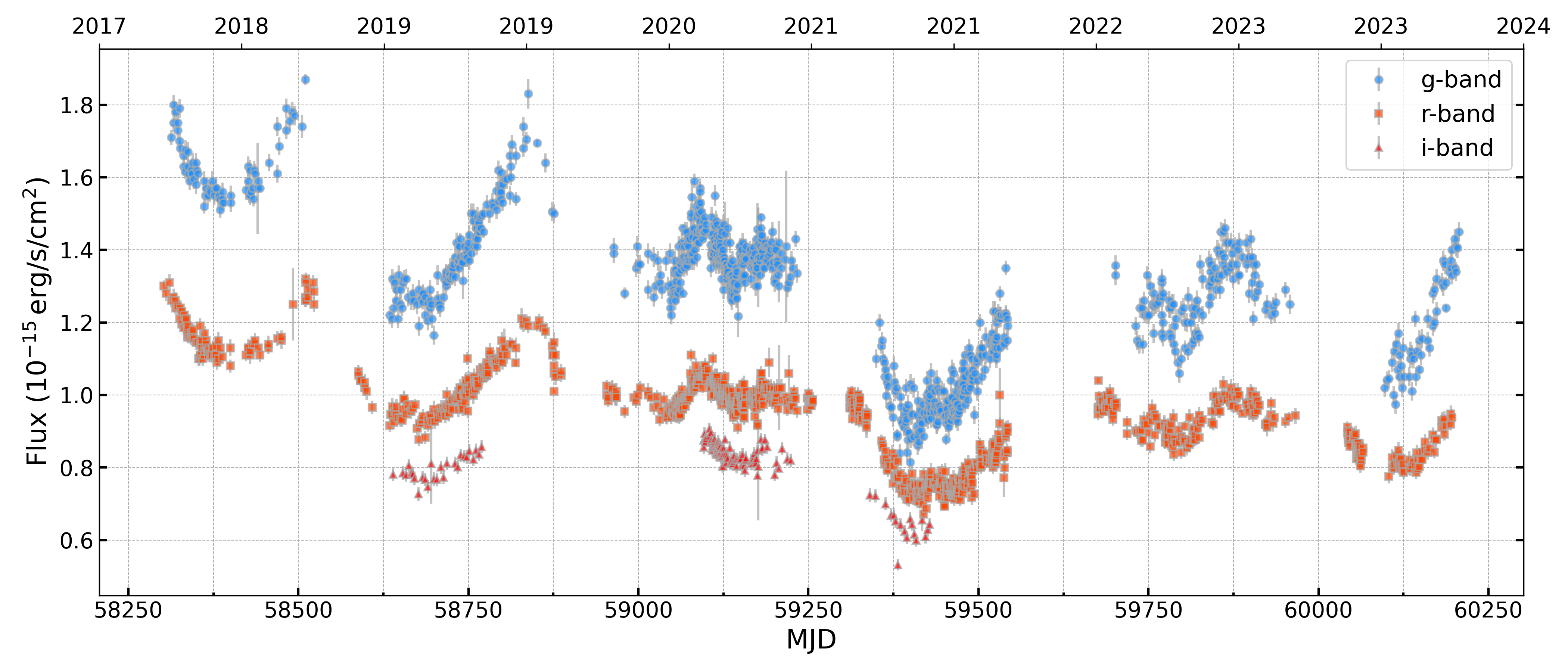}
        \caption{ZTF lightcurves in the three ZTF bands (g-, r-, and i-bands) for one of the AGNs (SDSS J225307.36+194234.6) from our filtered sample.}
        \label{fig:ztf_lc}
    \end{figure*}

We used the ZTF's forced photometry service \citep[see][]{masci2018zwicky, ztffp} and focused our analysis on the variability observed in the ZTF-g band. This band was chosen due to its broader coverage -- more data available for this band than for the others, and because it typically shows greater variability in AGNs compared to other bands \citep{cristiani1996optical, giveon1999long, helfand2001long}. It is important to note that the observer’s frame g-band samples different parts of the spectrum at varying redshifts. From the original sample of 2,056 quasars, we obtained at least one light curve for 1,527 using the ZTF dataset. After applying a quality filter, as described in Section \ref{sec:metho}, our final sample was refined to 915 quasars. For this final sample, we obtained light curves in the g and r bands, each with approximately 2000 Modified Julian Dates (MJDs) in the rest frame per source. An example of a ZTF light curve for one of our selected AGNs is shown in Figure \ref{fig:ztf_lc}.

\section{Methodology} \label{sec:metho} 

\subsection{Data retrieval and filtering process} \label{sec:data_ret}

Our data filtering methodology, implemented through a custom Python script \citep{Benati_Goncalves_ZTF-Data-Filters_2024}, is designed to improve the robustness of light curves from the ZTF dataset. Initially, we retrieve the requested data from the ZTF's forced photometry service using a simple Python script. We then begin filtering by removing ``NaN" values and resampling the light curves into one-day bin averages. We chose this binning method because our focus is on variations longer than one day, rather than intra-day variability, as implemented in previous studies that also utilized ZTF data \citep[e.g.,][]{ma2024h}.


Subsequently, employing statistical techniques, such as computing the mean and root mean square deviation (RMS) of flux values, the code identifies and filters out outliers within the light curves. Data points exceeding a threshold of four times the RMS deviation from a 30-day rolling mean are identified as potential outliers and excluded from the dataset, as this threshold is chosen because it encompasses approximately 99.99\% of the data in a normal distribution, effectively capturing nearly all legitimate observations while excluding extreme deviations that could be due to intra-night variations in the observation conditions, \textit{e.g.}, significant changes in airmass or cloud-cover between exposures.

Furthermore, we adopt a selection criterion requiring a minimum of 100 observations and a fractional variability ($F_{\text{var}}$) -- defined below -- to be greater than zero. $F_{\text{var}}$ is a well-established metric in variability studies \citep{Peterson_1998,VandenBerk_2004,Kaspi_2007, kozlowski_2010} and provides a normalized measure of intrinsic variability after accounting for measurement uncertainties. This parameter ensures that the variability detected in the light curves is intrinsic to the source and not dominated by noise or errors. Mathematically, $F_{\text{var}}$ is defined as:

\begin{equation}
    F_{\text{var}} = \sqrt{\frac{\sigma_{\text{XS}}^2}{\langle F \rangle^2}},
\end{equation}

where $\sigma_{\text{XS}}^2$ is the excess variance, defined as:

\begin{equation}
    \sigma_{\text{XS}}^2 = S^2 - \langle \sigma_{\text{err}}^2 \rangle,
\end{equation}

and

\begin{equation}
    S^2 = \frac{1}{N-1} \sum_{i=1}^{N} \left(F_i - \langle F \rangle\right)^2.
\end{equation}

\noindent, where $F_i$ represents the observed flux, $\langle F \rangle$, is the mean flux, N is the number of data points, $\sigma_{\text{err}}^2$ is the measurement error variance for each observation, and $\langle \sigma_{\text{err}}^2 \rangle$ is the average of the error variances over all observations. The uncertainty in $F_{\text{var}}$, denoted as $\Delta F_{\text{var}}$, is computed using:

\begin{equation}
    \Delta F_{\text{var}} = \sqrt{\left( \frac{\langle \sigma_{\text{err}}^2 \rangle}{\langle F \rangle^2} \right) + \left( \frac{\sigma_{\text{err}}}{\langle F \rangle} \right)^2}
\end{equation}

This fractional variability metric serves as a robust measure of the intrinsic variability of light curves and accounts for both the source's variability and the observational uncertainties. By using $F_{\text{var}}$, the variability is normalized by the mean flux, ensuring that there is no bias introduced by the distance of the sources. This allows for consistent measurement of intrinsic variability across all objects.

According to our selection criteria, light curves with $F_{\text{var}} < 0$ are excluded from the analysis as they do not exhibit significant intrinsic variability. This refined approach ensures that only the most reliable and informative data are retained for further analysis. The flowchart below illustrates the logical steps applied by the code, which is openly accessible on GitHub\footnote{\url{https://github.com/hbgon/ZTF-Data-Filters}} with reproducible examples and can be utilized by the community.

\begin{figure*}[!ht]
    \centering
    \includegraphics[width=\textwidth]{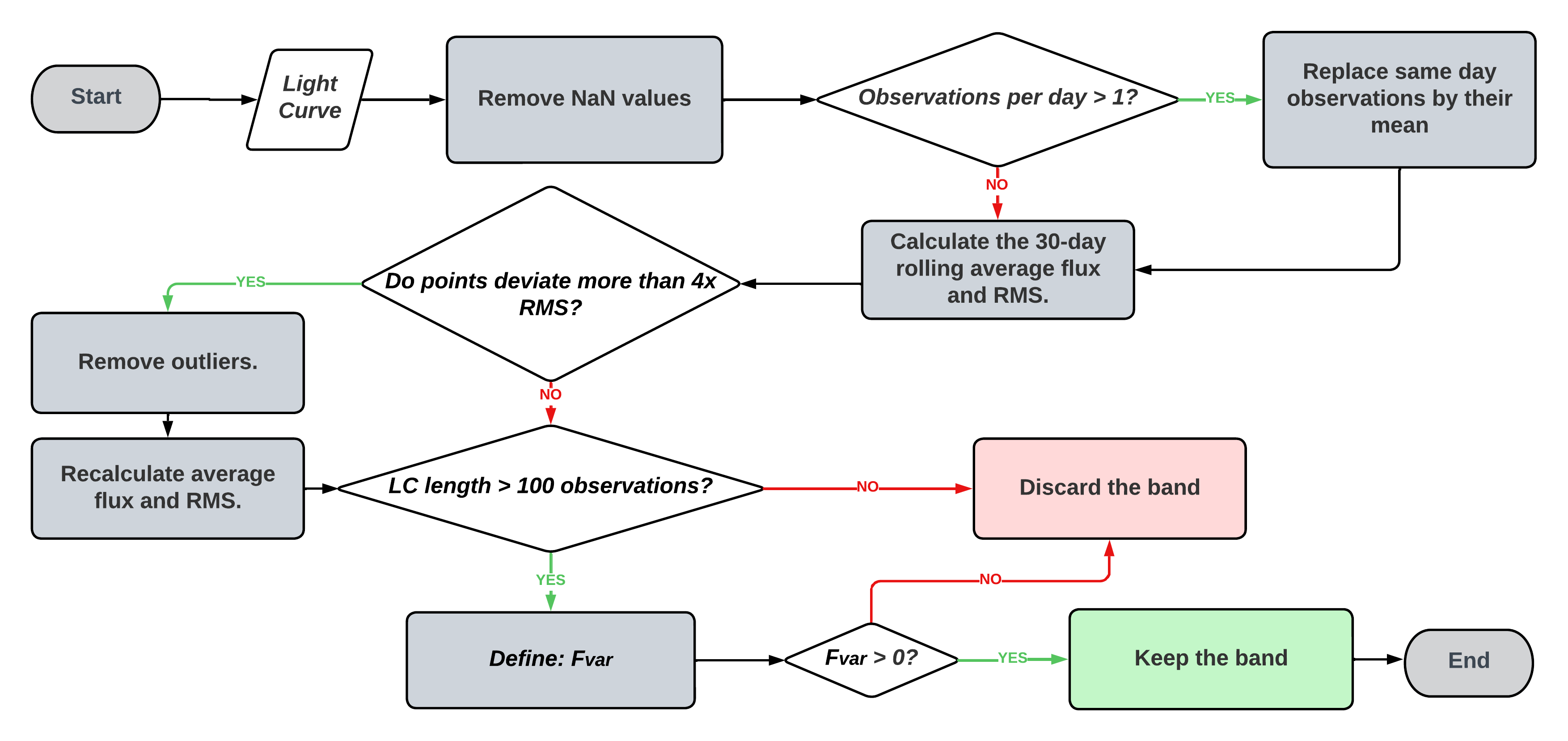}
    \caption{Flowchart illustrating the applied logical sequence for data filtering.}
    \label{flowchart}
\end{figure*}

\subsection{Emission line contribution to the photometric bands}
\label{sec:linecontributions}

To investigate the contribution of emission lines to the different filters, we selected the four quasar spectra with the best signal-to-noise ratio and the strongest emission lines from each redshift bin. One representative spectrum from each redshift bin is shown in Figure~\ref{fig:ztf+spec}, along with the identification of the main emission lines and the transmission curves of the ZTF filters. In different redshift bins, we observe varying contributions of emission lines within each ZTF filter. To ensure that our variability approach is not biased by the variation in these emission lines — meaning that some emission lines may contribute more to the filter than the continuum itself — we calculated the average contribution fraction of the emission lines in each filter for each redshift bin. This was done using emission line fluxes from the SDSS DR16 quasar catalog \citep{wu2022catalog}. Additionally, the fluxes of some emission lines were recalculated using PyQSOFit \citep{guo2018pyqsofit}, as their profiles were not fully covered by the filter's passband. In these cases, the fitted line profiles were convolved with the filter’s transmission curve to accurately estimate their contributions to the flux within the filter. 
In Table \ref{tab1}, we present the key findings, specifically the percentage contribution of each line to the continuum flux within the filter. The mean contributions of the line fluxes, relative to the continuum fluxes, are typically a few percent. Given that the fluxes of these lines generally vary by a certain percentage, their contribution to the overall variation is significantly lower compared to that of the continuum. Therefore, we have decided to analyze the light curves without adjusting for flux contributions from these lines.

To identify possible emission lines that may be included within each filter band at different redshifts, Table~\ref{tab:filter_wavelengths} lists the rest-frame wavelength coverage of the ZTF $g$, $r$, and $i$ filters across the redshift bins. This highlights how the filters shift with redshift and allows the verification of possible inclusion or absence of specific emission lines listed in Table~\ref{tab1}.

\begin{table*}
\centering
\scriptsize
\caption{Percent contributions of emission line fluxes within ZTF filters for different redshift bins, with the total mean percent contribution to the continuum fluxes (Cont.) given in the central column.}
\label{tab1}
\begin{tabular}{lll}
\toprule
\textbf{Filter} & \textbf{Cont.(\%)} & \textbf{Significant Lines (\%)} \\ \midrule
& & \textbf{0 $<$ z $<$ 0.25} \\
\hline
ZTF$_{\rm g}$ & 6.04 & [Ne~V] 3426 (0.42), H$\beta$ (2.07), He~II 4687 (0.83), H$\gamma$ (0.39), [O~III] 4959 (0.15),[O~III] 5007 (1.30), H$\delta$ (0.23), \\
& & [Ne~III] 3869 (0.24), [O~II] 3728 (0.41) \\
ZTF$_{\rm r}$ & 5.12 & [Ne~V] 3426 (0.55), H$\beta$ (1.87), He~II 4687 (0.71), H$\gamma$ (0.29)  \\ 
ZTF$_{\rm i}$ & 3.98 & [Ne~V] 3426 (0.60), H$\beta$ (1.77), He~II 4687 (0.78), H$\gamma$ (0.33)  \\ 
\hline
& & \textbf{0.25 $<$ z $<$ 0.75} \\
\hline
ZTF$_{\rm g}$ & 4.23 & Mg~II (4.12), [Ne~V] 3426 (0.11)  \\
ZTF$_{\rm r}$ & 1.48 & [Ne~V] 3426 (0.37), [O~II] 3728 (0.13), [Ne~III] 3869 (0.44), H$\gamma$ (0.33), H$\delta$ (0.21) \\
ZTF$_{\rm i}$ & 10.01 & H$\beta$ (5.84), He~II 4687 (0.64), H$\gamma$ (2.12), [O~III] 4959 (0.85), [O~III] 5007 (0.253), H$\delta$ (0.31) \\ 
\hline

& & \textbf{0.75 $<$ z $<$ 1.25} \\
\hline
ZTF$_{\rm g}$ & 2.83 & C~III (2.14), Mg~II (0.06), Al~III 1857 (0.34), Si~III 1892 (0.29)  \\
ZTF$_{\rm r}$ & 5.48 & Mg~II (5.34), [Ne~V] 3426 (0.14) \\
ZTF$_{\rm i}$ & 3.37 & [O~II] 3728 (0.19), H$\delta$ (1.01), [Ne~III] 3869 (0.92), H$\gamma$ (0.93), [Ne~V] 3426 (0.32) \\ 
\hline

& & \textbf{1.25 $<$ z $<$ 1.75} \\
\hline
ZTF$_{\rm g}$ & 0.36 & C~IV ($<$0.01\%), C~III (0.15), Al~III 1857 (0.04), Si~III 1892 (0.08), He~III 1640 (0.08) \\
ZTF$_{\rm r}$ & 1.43 & Mg~II (1.43) \\
ZTF$_{\rm i}$ & 4.05 & Mg~II (3.75), [Ne~V] 3426 (0.30) \\ 
\hline

& & \textbf{1.75 $<$ z $<$ 2.0} \\
\hline
ZTF$_{\rm g}$ & $<$0.01 & C~IV ($<$0.01\%), Ly$\alpha$ ($<$0.01\%) \\
ZTF$_{\rm r}$ & 2.88 & C~III (1.80), Al~III 1857 (0.43), Si~III 1892 (0.51), Ni~III 1750 (0.14) \\
ZTF$_{\rm i}$ & 4.20 & Mg~II (4.20),  \\ 
\hline

& & \textbf{2.0 $<$ z $<$ 2.5} \\
\hline
ZTF$_{\rm g}$ & $<$0.01 & Ly$\alpha$ ($<$0.01\%), C~IV ($<$0.01\%) \\
ZTF$_{\rm r}$ & 1.04 & C~III (0.16), C~IV (0.54), He~III 1640 (0.10), Al~III 1857 (0.09), Si~III 1892 (0.145) \\
ZTF$_{\rm i}$ & \text{None} & \text{None} \\ 
\hline

& & \textbf{2.5 $<$ z $<$ 3.0} \\
\hline
ZTF$_{\rm g}$ & 5.81 & Ly$\alpha$ (5.81) \\
ZTF$_{\rm r}$ & $<$0.01 & C~IV ($<$0.01\%) \\
ZTF$_{\rm i}$ & 1.74 & C~III (1.20), He~III 1640 (0.06), Al~III 1857 (0.24), Si~III 1892 (0.24) \\ 
\hline

\end{tabular}
\end{table*}

\begin{table}
\centering
\caption{Observed minimum and maximum wavelengths for ZTF g, r, and i filters across different redshift bins.}
\label{tab:filter_wavelengths}
\begin{tabular}{|c|c|c|c|}
\hline
\textbf{Filter} & \textbf{Redshift Bin} & \textbf{$\lambda_{\text{min}}$ (\AA)} & \textbf{$\lambda_{\text{max}}$ (\AA)} \\ \hline
\multirow{7}{*}{g} & $0 \leq z < 0.25$ & 3280 & 5500 \\ \cline{2-4} 
                   & $0.25 \leq z < 0.75$ & 2343 & 4400 \\ \cline{2-4}
                   & $0.75 \leq z < 1.25$ & 1754 & 3667 \\ \cline{2-4}
                   & $1.25 \leq z < 1.75$ & 1367 & 3056 \\ \cline{2-4}
                   & $1.75 \leq z < 2.0$ & 1166 & 2750 \\ \cline{2-4}
                   & $2.0 \leq z < 2.5$ & 1025 & 2444 \\ \cline{2-4}
                   & $2.5 \leq z \leq 3.0$ & 912 & 2200 \\ \hline
\multirow{7}{*}{r} & $0 \leq z < 0.25$ & 4560 & 7000 \\ \cline{2-4} 
                   & $0.25 \leq z < 0.75$ & 3429 & 5600 \\ \cline{2-4}
                   & $0.75 \leq z < 1.25$ & 2857 & 4667 \\ \cline{2-4}
                   & $1.25 \leq z < 1.75$ & 2375 & 3889 \\ \cline{2-4}
                   & $1.75 \leq z < 2.0$ & 2133 & 3500 \\ \cline{2-4}
                   & $2.0 \leq z < 2.5$ & 1900 & 3111 \\ \cline{2-4}
                   & $2.5 \leq z \leq 3.0$ & 1714 & 2800 \\ \hline
\multirow{7}{*}{i} & $0 \leq z < 0.25$ & 5600 & 8900 \\ \cline{2-4} 
                   & $0.25 \leq z < 0.75$ & 4214 & 7112 \\ \cline{2-4}
                   & $0.75 \leq z < 1.25$ & 3500 & 5922 \\ \cline{2-4}
                   & $1.25 \leq z < 1.75$ & 2929 & 4944 \\ \cline{2-4}
                   & $1.75 \leq z < 2.0$ & 2633 & 4450 \\ \cline{2-4}
                   & $2.0 \leq z < 2.5$ & 2333 & 3956 \\ \cline{2-4}
                   & $2.5 \leq z \leq 3.0$ & 2100 & 3556 \\ \hline
\end{tabular}
\end{table}
\section{Results and discussion}
\label{sec:results}

This section explores the variability of our sample, characterized by the $F_{\text{var}}$ of the ZTF g-band light curves. This selection is based on the observation that the g-band typically exhibits the highest variability, aligning with well-established research indicating that AGN continuum variations are more pronounced at shorter, bluer wavelengths \citep{Wandel_1999, Vaughan_2003, Peterson_2014, MacLeod2010}.

We investigate the relationship between variations in the g-band and the properties of quasars from the SDSS DR16 quasar catalog \citep{wu2022catalog}. We begin by addressing potential biases introduced by sample selection and observational strategies, with particular attention to the degeneracy between luminosity, black hole mass, redshift, and the corresponding central rest-frame wavelength.

\subsection{Potential biases}
\label{potential_biases}

\begin{figure*}[ht!]
    \centering
    \hspace{-1cm}
    \includegraphics[width=1.05\textwidth]{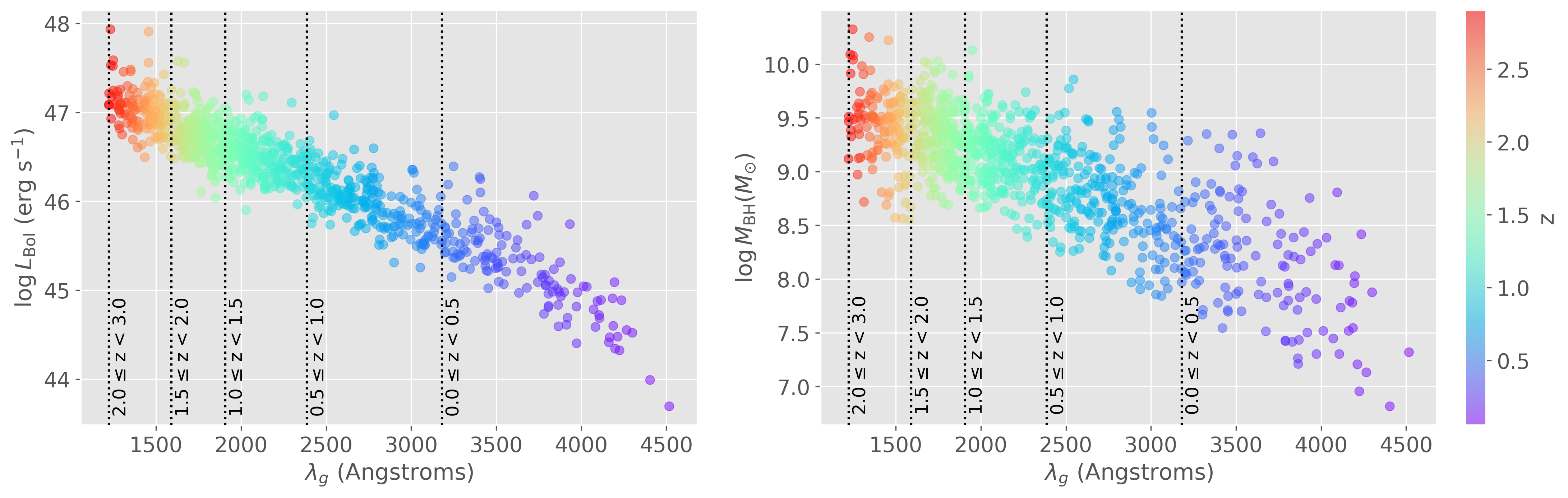}
    \caption{Relationship between bolometric luminosity (left) and black hole mass (right) with the central rest-frame wavelength at each quasar corresponding to the observed g-band central wavelength. The color gradient indicates varying redshifts, highlighting a decrease of both L$_{bol}$ and M$_{BH}$ with decreasing redshift and increasing wavelength. Dashed vertical lines delimit the wavelength ranges  corresponding to the redshift intervals indicated in the Figure.}
    \label{lbol_bhm_vs_lambda}
\end{figure*}

The analysis presented in this work can be biased in several ways. As shown in Figure \ref{lbol_bhm_vs_lambda}, both bolometric luminosity ($L_{\text{bol}}$) and black hole mass ($M_{\text{BH}}$) decrease with decreasing redshift and increasing wavelength. These relationships complicate a straightforward analysis of the photometric variability, measured via $F_{\text{var}}$, in relation to quasar properties, as they introduce biases when comparing these quantities across different redshift bins.

Our sample spans redshifts in the range $0\leq z<3$. Given that quasars at $2 < z < 3$ -- the so-called cosmic noon -- are, on average, more luminous than those at lower redshifts \citep[e.g.][]{Madau_2014}, the relationships between L$_{bol}$ and M$_{BH}$ with redshift and central rest-frame wavelength reveal the possible presence of biases in the investigation of the relation between F$_{\text{var}}$ and quasar properties.

In particular, our data are subject to at least three observational biases: (1) the higher the redshift, the larger the volume we probe, which contains a greater fraction of luminous sources; (2) the higher the redshift, the fainter a source with a given luminosity becomes, which also results in detecting a larger fraction of luminous sources, as the least luminous ones may become undetectable  -- the Malmquist bias \citep{Bigot1990, Butkevich2005}; (3) the length of the light curves in the rest frame varies with redshift. For nearby sources, the observed time interval covered by the data is approximately 2000 days, while for sources at increasing redshifts, the observed interval is reduced by a factor of \(1 + z\), thus the intervals get progressively shorter, such that at the highest redshifts of \(z \sim 3\), the observed rest-frame time interval gets down to approximately 500 days.

Given these complexities, to minimize the dependence on redshift we have segmented our sample into the following redshift bins: 0--0.5, 0.5--1, 1--1.5, 1.5--2, and 2--3. The bins were selected to separate the sources in redshift bins but at the same time ensure a sufficient number of objects in each bin, allowing for statistically meaningful results.

We present the first results of our analysis in histograms of $F_{\text{var}}$ values, separated into the redshift bins mentioned above in Figure \ref{$F_{var}$_gr}. The blue histogram represents $F_{\text{var}}$ values for the g-band, while the red histograms correspond to the r-band. As discussed earlier, $F_{\text{var}}$ for the g-band reaches higher values than for the r-band, supporting the use of $F_{\text{var,g}}$ as the best tracer of variability. The median values of $F_{\text{var,g}}$ range from 0.09 to 0.13, with the highest median value corresponding to the lowest redshift bin and the lowest median to the highest redshift bin. This result appears to contradict previous studies, which found that AGN continuum variations are more pronounced at shorter, bluer wavelengths \citep{Wandel_1999, Vaughan_2003, Peterson_2014, MacLeod2010}. However, this can be explained by the competing effect introduced by the fact that more luminous quasars tend to exhibit lower variability amplitudes \citep[e.g.][]{Chanchaiworawit_2024}, as will be discussed further in the next section.

\begin{figure*}[ht!]
    \centering
    \includegraphics[width=\textwidth]{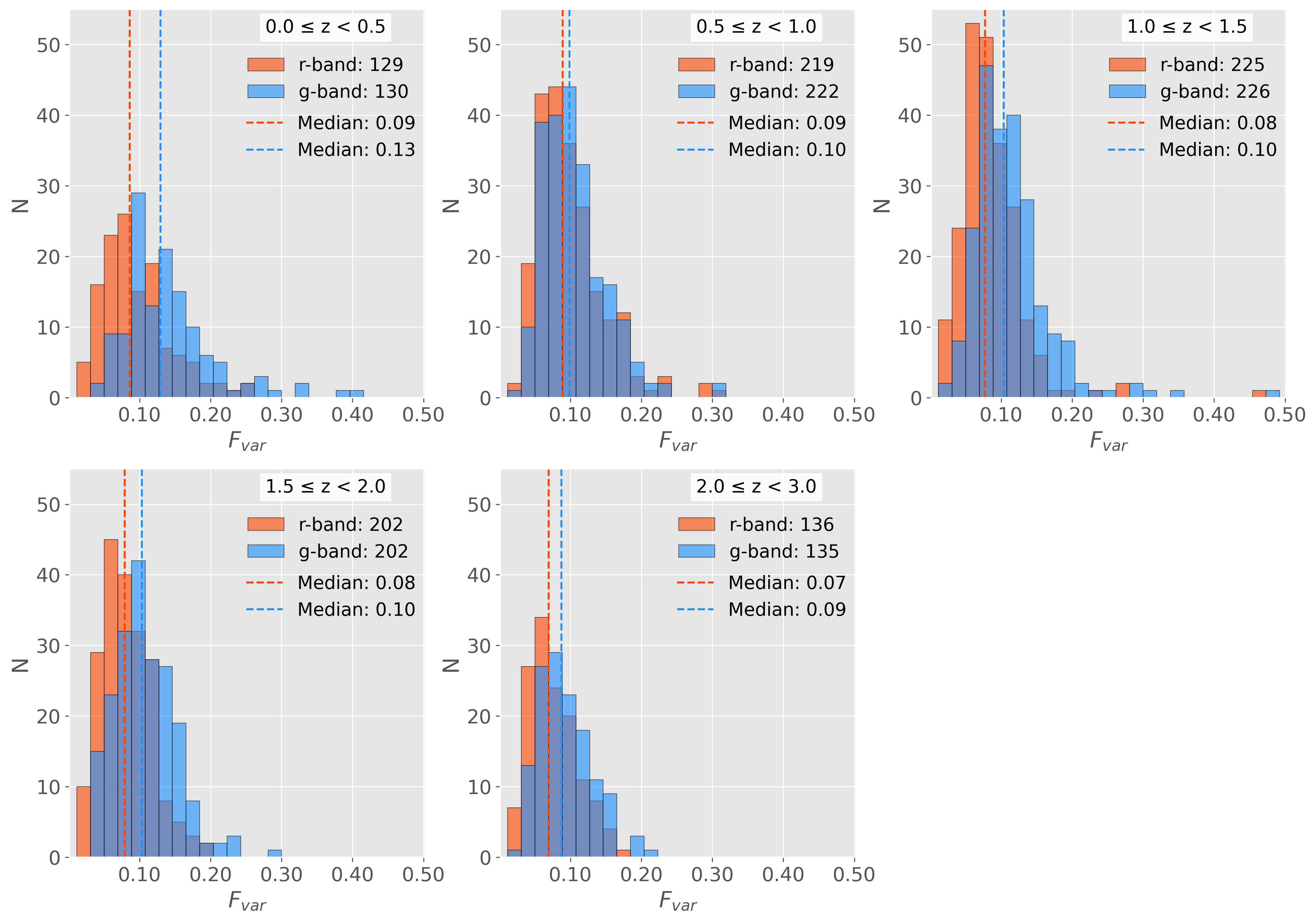}
    \caption{Histograms of $F_{\text{var}}$ for the sample in five redshift bins, showing both g- and r-band distributions. The total number of sources is given in the top right corner of each subplot.}
    \label{$F_{var}$_gr}
\end{figure*}

In the following sections, we investigate the relationships between quasar luminosities, black hole mass, Eddington ratio, and $F_{\text{var,g}}$ (hereafter referred to simply as $F_{\text{var}}$) across the different redshift bins mentioned above. In the context of statistical analysis, we assess the significance of observed trends by testing the null hypothesis that there is no correlation between the variables. A p-value threshold of 0.05 or lower is considered indicative of statistically significant correlations. Values slightly above this threshold (e.g., $p \leq 0.06$) are considered marginally significant, suggesting trends that may warrant further investigation and possibly reflect the complex interplay of factors we have identified.

To obtain linear regressions for each redshift bin, we employ the Weighted Least Squares (WLS) method, using intrinsic measurement errors in the spectral properties \citep{wu2022catalog} as weights. This ensures that data points with smaller uncertainties have a greater influence on the analysis. This approach improves the accuracy of our regression, particularly in handling the heteroscedastic nature of astronomical data. The resulting p-value and Pearson's correlation coefficient ($r$) for each bin are shown with the regression fits in distinct colors. These are highlighted in Figures~\ref{$F_{var}$_l_binned}, \ref{mass_$F_{var}$_redshift_binned}, and \ref{eddington_$F_{var}$_binned} in the following sections.

\subsection{Luminosity vs. \texorpdfstring{$F_{\mathrm{var}}$}{Fvar}}
\label{sec:l_vs_fvar}

Figure \ref{$F_{var}$_l_binned} explores the relationship between photometric variability ($F_{\text{var}}$) and four key luminosities: $L_{1350}$, $L_{3000}$, $L_{5100}$, and bolometric luminosity. These specific wavelengths were chosen to probe distinct regions of the quasar accretion disk and the continuum emission emanating from it. $L_{1350}$ and $L_{3000}$ trace the ultraviolet continuum, predominantly sensitive to the innermost disk regions \citep{Richards2006, Shen2011, Panda_2024_virialmasses}, while $L_{5100}$ reflects the optical continuum, associated with extended disk regions \citep{VandenBerk2004}. Bolometric luminosity provides an integrated view of the total radiative output, encompassing multiple emission components \citep[see e.g.,][]{Netzer_2019}. The luminosity derived from the [O III] $\lambda 5007$ emission line was also considered. However, no significant correlation with $F_{\text{var}}$ was found, as expected given the long timescales over which variability in the narrow-line region should occur.

The absence of low-redshift bins for $L_{1350}$ and high-redshift bins for $L_{5100}$ arises from the SDSS spectral coverage: at low redshifts, $L_{1350}$ lies outside the observed range, while at high redshifts, $L_{5100}$ is redshifted beyond the SDSS wavelength limit. This restricts the redshift intervals in which these luminosities can be measured.

A notable segregation of points is seen for $L_{3000}$, $L_{5100}$, and bolometric luminosities as a function of of redshift, with sources at higher redshifts showing the highest luminosities. This pattern is influenced not only by selection biases, as quasars at higher redshifts must be sufficiently luminous to be detectable \citep{Richards2006, Panda_2024_virialmasses}-- see also the discussion in Sec.\,\ref{potential_biases}. We cannot also discard the potential intrinsic evolutionary changes in quasar properties over cosmic time \citep{MacLeod2010}. The interplay between these factors might obscure underlying trends by preferentially sampling more luminous quasars at higher redshifts, complicating our understanding of their inherent variability.

A general trend seen in Fig. \ref{$F_{var}$_l_binned} is an anti-correlation between the luminosities and $F_{\text{var}}$, seen for all luminosities and across all redshift bins, in agreement with results of previous studies \citep{Chanchaiworawit_2024}, with the strongest correlations been observed at the highest redshifts.

\begin{figure*}[!ht]
    \centering
    \includegraphics[width=\textwidth]{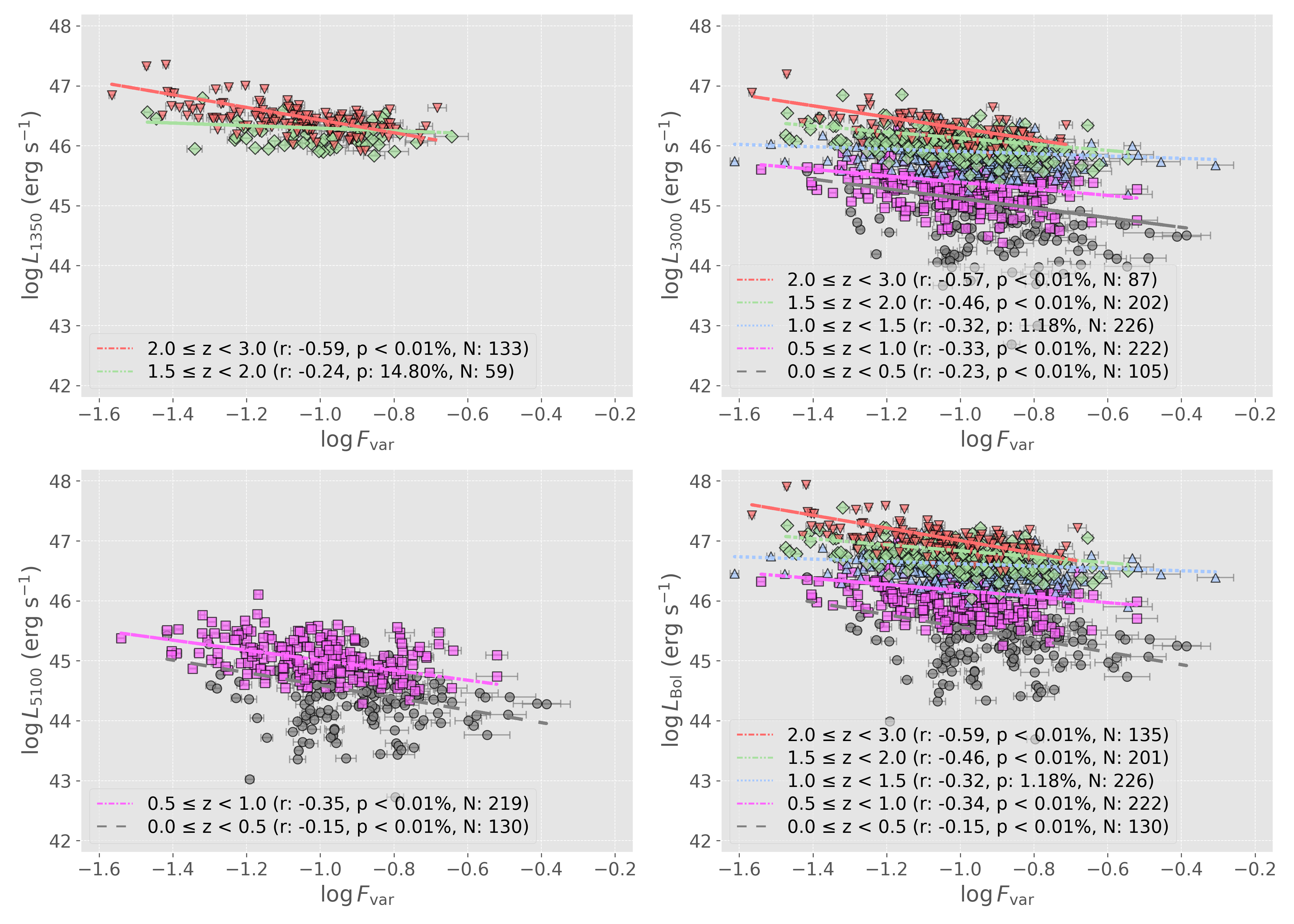}
    \caption{Luminosities $L_{1350}$, $L_{3000}$, $L_{5100}$, and bolometric luminosity versus $F_{\text{var}}$ in different redshift bins, with linear regression fits overlaid. For the highest (lowest) redshift bins, the wavelength range of the spectra did no cover L$_{5100}$ (L$_{1350}$).  Pearson's correlation coefficient (r), p-value (p), and total number of points (\(N\)) are provided at the bottom of each plot.}
    \label{$F_{var}$_l_binned}
\end{figure*}

Although the ZTF $g$-band filter spans approximately 4000\AA\ to 5500\AA\ at z$\approx$0, but bluer spectral regions at higher z, and does not directly encompass the selected luminosities at higher redshifts, its variability still probes the variation of the AGN continuum that extends over the whole wavelength range. Previous studies have shown that variability in the optical continuum often correlates with changes in the accretion disk and the reprocessing of high-energy radiation \citep{Kelly2009, MacLeod2010}. This makes analyzing $F_{\text{var}}$ in the $g$-band a robust approach, even when the covered wavelength region of the sources extends to much bluer wavelengths than those at z$\approx$0 as the redshifts become higher.


\subsection{Black hole mass vs. \texorpdfstring{$F_{\mathrm{var}}$}{Fvar}}
\label{sec:bhm_vs_fvar}

In this section, we explore the relationship between black hole mass ($M_{\mathrm{BH}}$) and $F_{\text{var}}$ as shown in Figure\,\ref{mass_$F_{var}$_redshift_binned}, where the data points are colored by redshift. A clear segregation is observed, with distinct trends across different redshift ranges. However, this apparent segregation needs to be carefully considered, as it could be influenced by a combination of observational biases and systematic effects.

One notable factor is the potential impact of luminosity bias, also present in the analysis of Section \ref{sec:l_vs_fvar}. Additionally, the methods used to estimate black hole masses introduce further complexity. The cataloged masses, derived from broad emission line widths, are subject to varying degrees of bias depending on the specific line used, as discussed in great detail in \citet{wu2022catalog}. For the redshift range considered here, the Mg II line is generally more reliable, as it is less affected by the systematic uncertainties that compromise C IV-based estimates, especially at higher redshifts. Moreover, differences in the Eddington ratio with redshift could influence the observed relationship, as changes in accretion efficiency may drive differences in photometric variability across cosmic time, adding another layer of complexity. These effects, combined with possible selection biases, underscore the need for caution when interpreting these results.

To examine this relationship in more detail, we again divided the sample into five redshift bins, as previously defined. The results of the analysis, including the linear regression fits and Pearson's correlation coefficient (r), are summarized in Figure \ref{mass_$F_{var}$_redshift_binned}. 

\begin{figure}[!ht]
    \centering
    \hspace{-1.0cm}
    \includegraphics[width=1.1\linewidth]{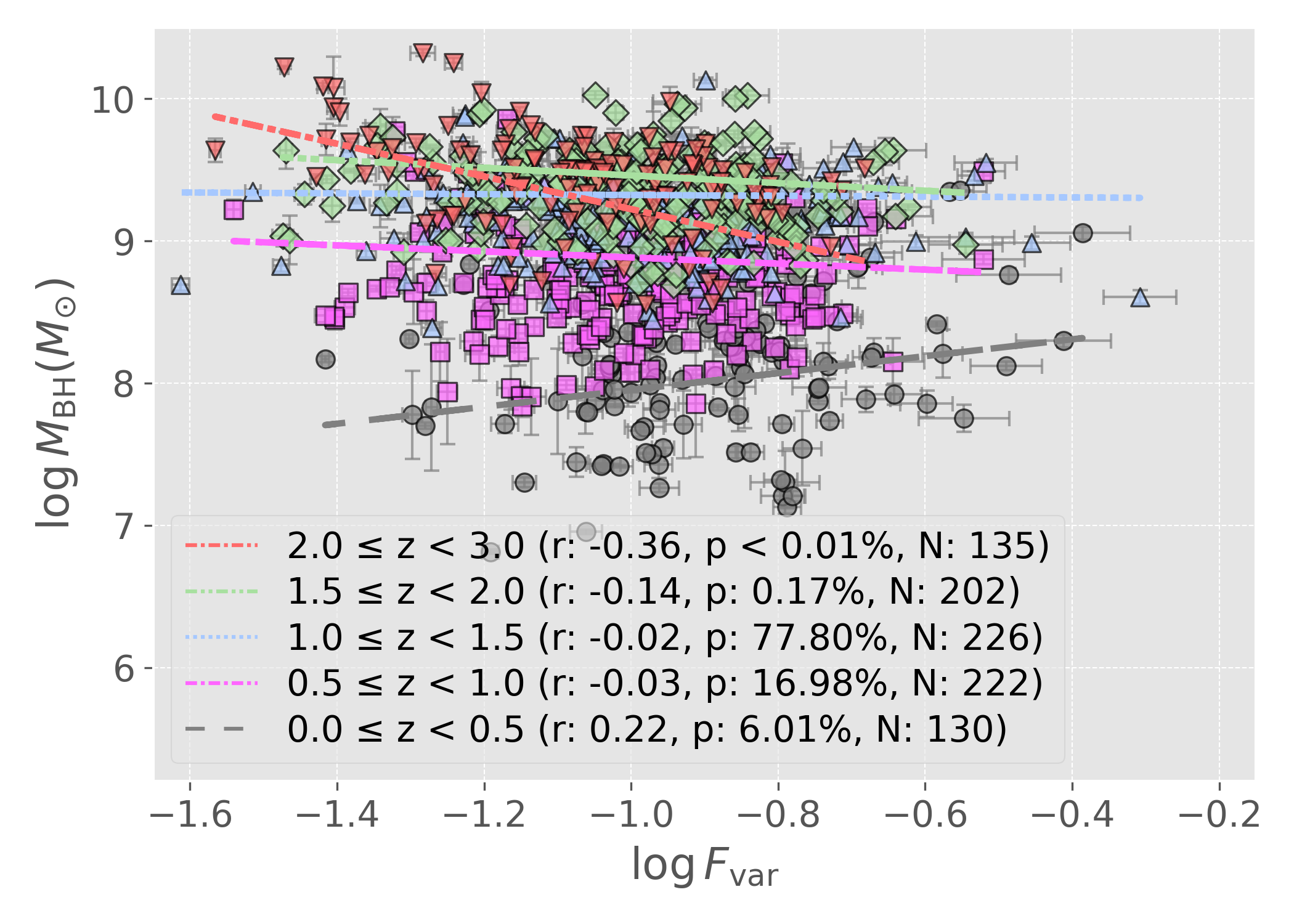}
    \caption{Black hole mass versus $F_{\text{var}}$ in different redshift bins, with linear regression fits overlaid. The Pearson's correlation coefficient (\(r\)), p-value (\(p\)), and the total number of points (\(N\)) are shown at the bottom.}
    \label{mass_$F_{var}$_redshift_binned}
\end{figure}


At low redshifts (\( z < 0.5 \)), the analysis reveals a weak positive correlation between $M_{\mathrm{BH}}$ and $F_{\text{var}}$, with a Pearson correlation coefficient of $r \sim 0.22$ and a p-value indicating marginal significance ($p \sim 6\%$). This trend aligns with findings from earlier works \citep{wold2007dependence,lu2019supermassive}, who observed similar behavior in local AGN samples. However, as redshift increases, the correlation weakens and eventually inverts. In the highest redshift bin (\( z > 2.0 \)), a pronounced anti-correlation emerges, with a strong negative Pearson's coefficient and a statistically significant p-value ($r \sim -0.37$, $p < 0.01\%$). This result is consistent with studies of high-redshift quasars \citep{Kelly2009,kelly2013active}, where similar trends have been reported. This anti-correlation may be related to the tumultuous conditions prevalent during ``cosmic noon", which characterize the environments of young quasars. During these formative periods, growing black holes undergo phases of vigorous accretion marked by high instability \citep{Fan2006, Volonteri2010, Treister2011}. Additionally, the presence of the Ly$\alpha$ forest in the $g$-band at this redshift could also contribute. The absorption lines affect the continuum flux and dilute the variability signal, with different quasars potentially experiencing varying degrees of this effect.


\subsection{Eddington Ratio vs. \texorpdfstring{$F_{\mathrm{var}}$}{Fvar}}

The Eddington ratio $\lambda_{\mathrm{Edd}}=L_{Edd}/L_{bol}$, as shown in Figure \ref{eddington_$F_{var}$_binned}, reveals an anti-correlation with $F_{\text{var}}$, which persists across all redshift  bins, with the WLS fitted to the data being all very similar for $0<z<2$, while the last redshift bin, $2<z<3$, shows somewhat different slope and intercept values. The best-fit parameters for each redshift bin, as well as for all redshifts combined, are summarized in Table~\ref{tab:fit_params}.

The relationship between $\lambda_{\mathrm{Edd}}$ and $F_{\text{var}}$ offers valuable insights into the accretion physics at play. Interestingly, the data reveal an inverse trend: quasars with higher Eddington ratios generally exhibit lower variability. 

This inverse relationship may be attributed to the increased radiative efficiency of accretion flows at higher $\lambda_{\mathrm{Edd}}$. As the accretion disk becomes more radiatively efficient, it likely leads to smoother energy output and suppresses stochastic variability. 

Since $\lambda_{\mathrm{Edd}}$ is a tracer of the mass accretion rate to the source, this can be understood as follows: when the accretion rate is high, and more mass is being accreted, a variation in this value has a lower relative amplitude compared to when the accretion rate is low. This observation aligns with theoretical predictions and observational trends noted in previous studies \citep{Kelly2009, Ai_2010, Yu_2022}. This result has also been explained in modeling by \citet{elitzur2014evolution} and in spectral observations of nearby AGN with broad double-peaked Balmer-line profiles, which tend to be less luminous and show higher variability than non-double-peaked sources \citep{storchi2017double}.

The utilization of $\lambda_{\mathrm{Edd}}$ also normalizes the bolometric luminosity ($L_{\text{bol}}$) by the Eddington luminosity and thus inherently reducing both mass and redshift dependencies. Consequently, the relationship between $\lambda_{\mathrm{Edd}}$ and $F_{\text{var}}$ becomes more robust and less dependent on redshift-related effects. However, some luminosity dependence should still be present due to the calculation of the SMBH mass via the virial hypothesis \citep{Shen2011}.

Eddington ratio seems to be the key parameter for many other AGN properties too; at low z, a recent study has shown that it is also related to the AGN obscuration \citep[e.g.,][]{Ananna2022}.

As indicated by Figure \ref{eddington_$F_{var}$_binned}, the highest redshift bin ($2.0 \leq z < 3.0$) exhibits a weaker anti-correlation with $F_{\text{var}}$ than the other redshift bins. This discrepancy could be attributed to selection effects, such as the preferential detection of brighter, more highly accreting quasars, which may all be less variable \citep[see e.g.,][]{Lu2019}. To address this, we tested the fit excluding the highest redshift bin, but the correlation coefficient remained unchanged.

\begin{figure}[!ht]
    \centering
    \hspace{-1.0cm}
    \includegraphics[width=1.1\linewidth]{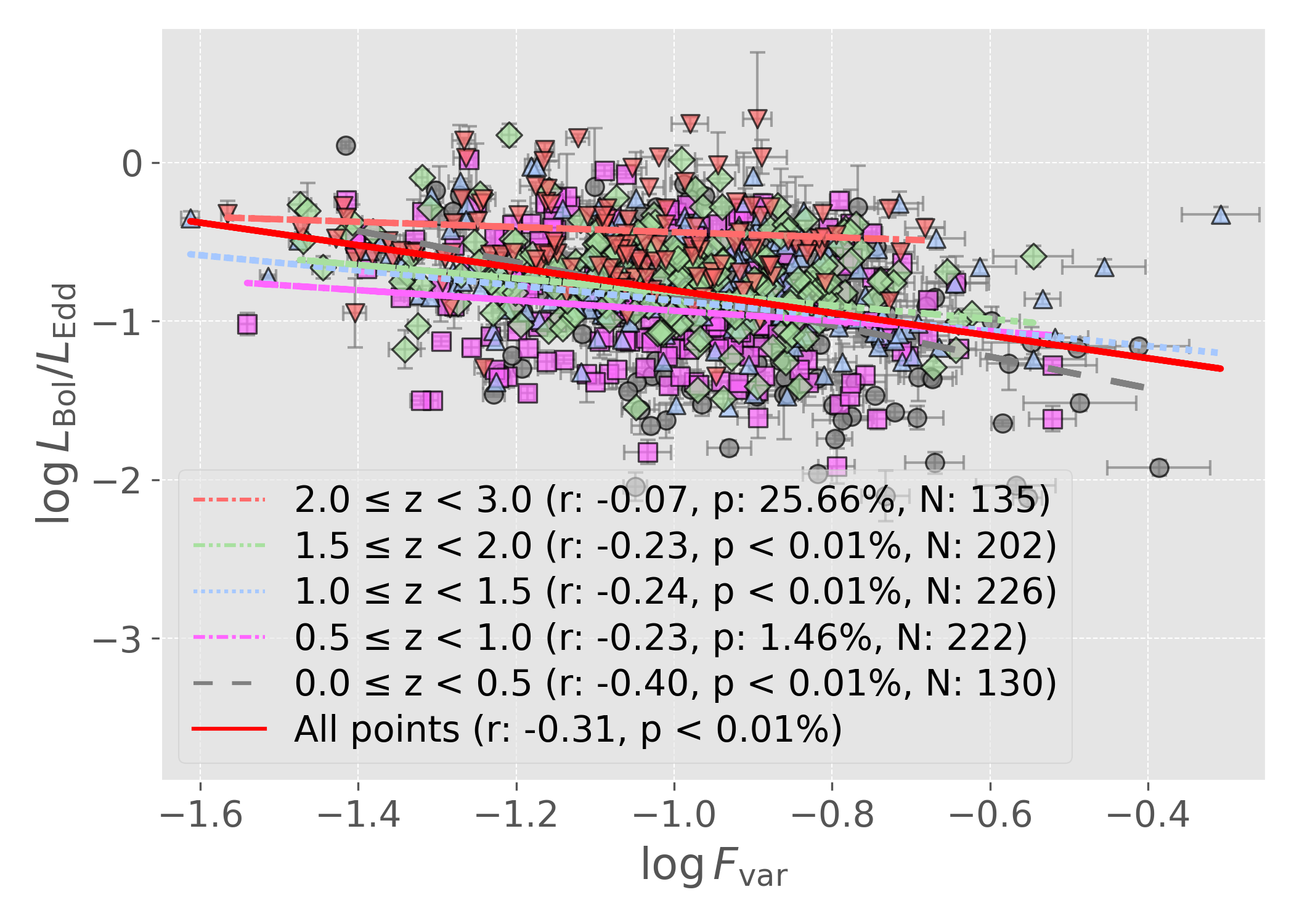}
    \caption{$\lambda_{\mathrm{Edd}}$ versus $F_{\text{var}}$ across different redshift bins, with linear regression fits overlaid. The Pearson's correlation coefficient (\(r\)), p-value (\(p\)), and the total number of points (\(N\)) are displayed at the bottom. Additionally, a fit for all points, along with its corresponding \(r\) and \(p\) values, is included.}
    \label{eddington_$F_{var}$_binned}
\end{figure}

\begin{table}
    \centering
    \caption{Best-fit parameters for the regression of $\lambda_{\mathrm{Edd}}$ versus \(F_{\mathrm{var}}\).}
    \label{tab:fit_params}
    \begin{tabular}{lccc}
        \hline
        \textbf{Redshift Bin} & \textbf{Slope (\(a\))} & \textbf{Intercept (\(b\))} & \(\mathbf{R^2}\) \\
        \hline
        \(0.0 \leq z < 0.5\) & -0.99 & -1.82 & 0.18 \\
        \(0.5 \leq z < 1.0\) & -0.32 & -1.26 & 0.03 \\
        \(1.0 \leq z < 1.5\) & -0.48 & -1.35 & 0.10 \\
        \(1.5 \leq z < 2.0\) & -0.43 & -1.24 & 0.12 \\
        \(2.0 \leq z < 3.0\) & -0.16 & -0.60 & 0.01 \\
        \textbf{Overall} & \textbf{-0.71} & \textbf{-1.52} & \textbf{0.15} \\
        \hline
    \end{tabular}
\end{table}

\begin{figure}[!ht]
    \centering
    \hspace{-1cm}
    \includegraphics[width=1.1\columnwidth]{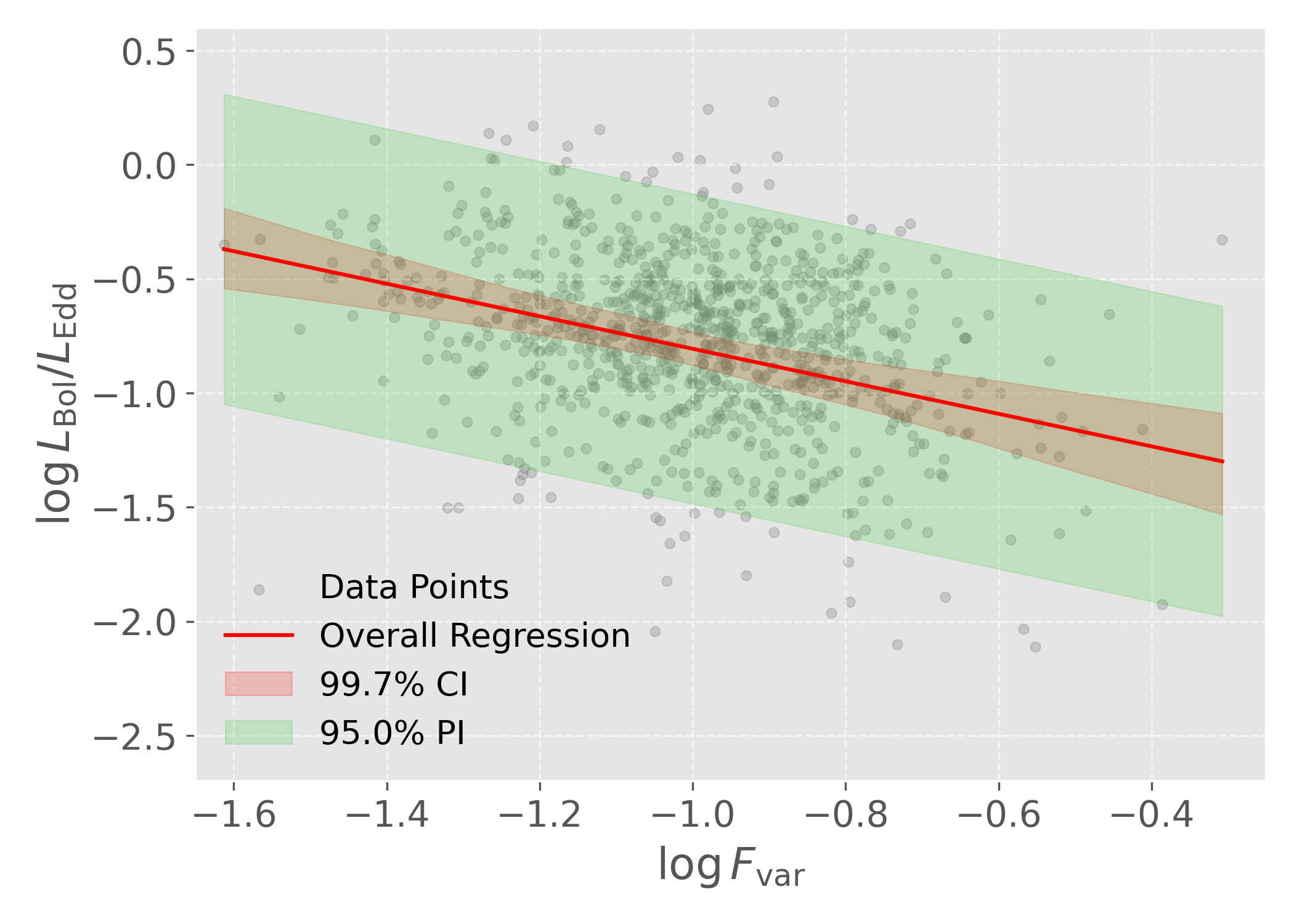}
    \caption{Same as Figure~\ref{eddington_$F_{var}$_binned}, but now showing gray circles representing all the data points (all redshifts). The overall regression is shown as a solid red line. The shaded red region represents the 99.7\% (3$\sigma$) confidence interval, and the green region indicates the 95\% (2$\sigma$) prediction interval. The slope, intercept, and $R^2$ value of the overall regression are displayed at the bottom-left corner.}
    \label{eddington_relation}
\end{figure}

To emphasize this relationship further, we present the entire dataset as a single ensemble and show the general relationship between the $\lambda_{\mathrm{Edd}}$ and $F_{\text{var}}$ in Fig \ref{eddington_relation}. The green-shaded region in Figure \ref{eddington_relation} represents the general regression's 95\% confidence interval. 

The combined regression of the entire sample, shown in Fig.\ref{eddington_relation} establishes a general relationship between $\lambda_{\mathrm{Edd}}$ and $F_{\text{var}}$. The derived equation, shown in Eq.~\ref{eq:general_relation}, captures the global anti-correlation observed: 

\begin{equation}
    \log \lambda_{\mathrm{Edd}} = (-0.71 \text{ ± } 0.06) \cdot \log F_{\mathrm{var}} - (1.52 \text{ ± } 0.06)
    \label{eq:general_relation}
\end{equation}

This relation is particularly useful to estimate the Eddington ratio of a general quasar at $0<z<3$ from its measured $F_{\text{var}}$ and provides a potential framework for exploring similar trends in other AGN populations and instruments. 


Nevertheless, we point out that it is important to explore further the validity of equation \ref{eq:general_relation} at higher redshifts than $\approx$\,2, with data obtained with more sensitive instruments -- such as those soon to be obtained with the Vera Rubin telescope for the LSST survey, considering the fact that the poorer correlation was for the highest redshift bin ($2\le z \le 3$). Another possibility is that the rest frame time interval corresponding to observations at $z \gtrsim 2$, which is shorter than $\approx 2$ years (given the $\approx 6$ years/(1+$z$) observed-frame baseline), may be insufficient to fully capture the variability of these sources, especially when compared to the longer time intervals probed for lower-redshift sources.



\section{Conclusions}
\label{sec:conclusion}

This work examined the photometric variability ($F_{\text{var}}$) of a sample of 915 quasars with redshifts $0\leq z<3$ and its correlation with key spectral properties. Our findings are summarized as follows:

\begin{itemize}

\item A strong anti-correlation was observed between $F_{\text{var}}$ and continuum luminosities (e.g., $L_{1350}$, $L_{3000}$, $L_{5100}$) across the quasar sample. As expected, the bolometric luminosity also exhibited a consistent anti-correlation with $F_{\text{var}}$ across all redshift bins, reaffirming the universality of this relationship.

\item A redshift-dependent relationship between $F_{\text{var}}$ and black hole mass was observed, transitioning from a weak positive correlation at low redshifts to a moderate ($r = -0.31$, $p < 0.01$) anti-correlation at higher redshifts.

\item A persistent anti-correlation was identified between the $\lambda_{\mathrm{Edd}}$ and $F_{\text{var}}$ across all redshifts, with a global correlation coefficient of $r = -0.31$ (with $p < 0.01$), and the strongest correlation in the lowest redshift bin ($r = -0.4$, $p < 0.01$). Using the entire sample, a general relationship was derived, providing a valuable basis for exploring and interpreting AGN variability.

\end{itemize}

Further work is needed to address potential selection biases and refine corrections for systematic effects, ensuring the robustness of the observed correlations between $F_{\text{var}}$ and key AGN properties. While the lack of significant redshift dependence in the $\lambda_{\mathrm{Edd}}$ vs. $F_{\text{var}}$ anti-correlation highlights its fundamental role in quasar variability, exploring the physical mechanisms underlying these trends remains a crucial avenue for future research. The advent of next-generation surveys, such as the Vera C. Rubin Observatory's LSST, will provide larger samples and higher-cadence observations, enabling a more comprehensive understanding of variability across a broader AGN population. These advancements promise not only to refine our theoretical models but also to enhance our understanding of the central engines of quasars.

\begin{acknowledgments}

This work is based on observations obtained with the Samuel Oschin Telescope 48-inch and the 60-inch Telescope at the Palomar Observatory as part of the Zwicky Transient Facility project. ZTF is supported by the National Science Foundation under Grant No. AST-1440341 and a collaboration including Caltech, IPAC, the Weizmann Institute of Science, the Oskar Klein Center at Stockholm University, the University of Maryland, the University of Washington, Deutsches Elektronen-Synchrotron and Humboldt University, Los Alamos National Laboratories, the TANGO Consortium of Taiwan, the University of Wisconsin at Milwaukee, and Lawrence Berkeley National Laboratories. We thank Muryel Guolo for his invaluable assistance in explaining the operation of the ZTF and providing the code to process the light curves. His support was crucial to the success of this project. HBG acknowledges the financial support of the Coordenação de Aperfeiçoamento de Pessoal de Nível Superior (CAPES) under the Finance Code 001. TSB acknowledges the financial support of the Conselho Nacional de Desenvolvimento Científico e Tecnológico (CNPq) and CAPES. SP is supported by the international Gemini Observatory, a program of NSF NOIRLab, which is managed by the Association of Universities for Research in Astronomy (AURA) under a cooperative agreement with the U.S. National Science Foundation, on behalf of the Gemini partnership of Argentina, Brazil, Canada, Chile, the Republic of Korea, and the United States of America. SP acknowledges the financial support of the Conselho Nacional de Desenvolvimento Científico e Tecnológico (CNPq) Fellowships 300936/2023-0 and 301628/2024-6.
\end{acknowledgments}

\newpage
\bibliography{ref}{}
\bibliographystyle{aasjournal}

\end{document}